\documentclass[11pt,onecolumn,draftcls]{IEEEtran}
\pdfoutput=1
\usepackage[cmex10]{amsmath}
\usepackage{amsfonts}
\usepackage{amsthm}
\usepackage{graphicx}
\usepackage{algorithmic}
\usepackage{algorithm}
\usepackage{url}
\usepackage{multirow}
\usepackage{colortbl}
\usepackage{cite}

\newcommand{ \vect}[1]{\mathbf{#1}}
\newcommand{ \set}[1]{\mathcal{#1}}
\newcommand{ \mat}[1]{\mathbf{#1}}

\newcommand{ \norm}[1]{\| #1 \|}
\newcommand{ \eq}[1]{\mathcal{#1}}

\newcommand{ \cond}[1]{\langle #1 \rangle}
\newcommand{ \kl}[2]{KL( #1 \, \| \, #2)}

\newcommand{ \bx}{{\vect{x}}}
\newcommand{ \bt}{{\vect{t}}}
\newcommand{ \bz}{{\vect{z}}}
\newcommand{ \bnull}{{\vect{0}}}

\newcommand{ \bI}{{\mat{I}}}
\newcommand{ \bW}{{\mat{W}}}
\newcommand{ \bC}{{\mat{C}}}
\newcommand{ \bM}{{\mat{M}}}
\newcommand{ \bS}{{\mat{S}}}
\newcommand{ \bQ}{{\mat{Q}}}

\newcommand{ \cX}{{\set{X}}}

\newcommand{ \cN}{\mathcal{N}}

\newcommand{ \reals}{{\mathbb{R}}}
\newcommand{ \E}{{\mathbb{E}}}

\newcommand{ \bmu}{\mbox{\boldmath$\mu$}}

\newcommand{ \tr}{\mbox{tr}}

\DeclareMathOperator*{\argmin}{arg\,min}
\DeclareMathOperator*{\argmax}{arg\,max}

\begin{document}
%
\title{Statistical File Matching of \\ Flow Cytometry Data}

%
%
%

\author{Gyemin~Lee, ~William~Finn,~and~Clayton~Scott
\thanks{
G. Lee and C. Scott are with the Department of Electrical Engineering and Computer Science, University of Michigan, Ann Arbor,
MI, USA. E-mail: \{gyemin, cscott\}@eecs.umich.edu.
W. Finn is with the Department of Pathology, University of Michigan, Ann Arbor, MI, USA. E-mail: wgfinn@umich.edu.}}

\maketitle

\begin{abstract}
\boldmath
Flow cytometry is a technology that rapidly measures antigen-based markers associated to cells in a cell population.  
Although analysis of flow cytometry data has traditionally considered one or two markers at a time, there has been increasing interest in multidimensional analysis.  
However, flow cytometers are limited in the number of markers they can jointly observe, which is typically a fraction of the number of markers of interest. 
For this reason, practitioners often perform multiple assays based on different, overlapping combinations of markers.  
In this paper, we address the challenge of imputing the high dimensional jointly distributed values of marker attributes based on overlapping marginal observations.  
We show that simple nearest neighbor based imputation can lead to spurious subpopulations in the imputed data, and introduce an alternative approach based on nearest neighbor imputation restricted to a cell's subpopulation.  
This requires us to perform clustering with missing data, which we address with a mixture model approach and novel EM algorithm.  
Since mixture model fitting may be ill-posed, we also develop techniques to initialize the EM algorithm using domain knowledge.  
We demonstrate our approach on real flow cytometry data.
\end{abstract}



%
\section{Introduction}
\label{sec:intro}

Flow cytometry is a technique for quantitative cell analysis \cite{shapiro94}. 
It provides simultaneous measurements of multiple characteristics of individual cells.  
Typically, a large number of cells are analyzed in a short period of time -- up to thousands of cells per second.  
Since its development in the late 1960s, flow cytometry has become an essential tool in various biological and medical laboratories.  
Major applications of flow cytometry include hematological immunophenotyping and diagnosis of diseases such as acute leukemias, chronic lymphoproliferative disorders, and malignant lymphomas \cite{brown00clichem}.

Flow cytometry data has traditionally been analyzed by visual inspection of one-dimensional histograms or two-dimensional scatter plots. 
Clinicians will visually inspect a sequence of scatter plots based on different pairwise marker combinations, and perform gating, the manual selection of marker thresholds, to eliminate certain subpopulations of cells.  
They identify various pathologies based on the shape of cell subpopulations in these scatter plots.  
There has been recent work, reviewed below, on automatic cell gating or classification of pathologies based on multidimensional analysis of flow cytometry data.

Despite the promise of multidimensional analysis, this direction is limited by the number of markers that can be simultaneously measured, which is typically a fraction of the number of markers of interest. 
It is therefore common in practice to perform multiple assays based on different, overlapping combinations of markers.  
We may view these combinations as different marginals of the joint distribution of all observed markers.  
However, even when analysis is based on visual inspection of scatter plots, problems arise when the desired marker pair was not jointly measured.  
This situation arises frequently in the analysis of historical data.

To address these issues and to facilitate higher dimensional analysis,
we present a statistical method for file matching, which imputes higher dimensional flow cytometry data from multiple lower dimensional data files.
While \cite{pedreira08cytometry} proposed a simple approach based on Nearest Neighbor (NN) imputation, this method is prone to induce spurious clusters, as we demonstrate below. 
Our method can improve the file matching of flow cytometry and is less likely to generate false clusters.

In the following, we explain the principles of flow cytometry and introduce the file matching problem in the context of flow cytometry data.
We then present an approach to file matching which imputes a cell's missing marker values with the values of the nearest neighbor among cells of the same type.  
To implement this approach we develop a method for clustering with missing data.  
We model flow cytometry data with a latent variable Gaussian mixture model, where each Gaussian corresponds to a cell type, and develop an expectation-maximization (EM) algorithm to fit the model.  
Since a large majority of all values are unobserved, most covariances cannot be estimated from the data.  
However, domain experts possess considerable knowledge about the characteristics of different cell types, and we incorporate this knowledge into the initialization of the EM algorithm.
We compare our method with simple nearest neighbor imputation on real flow cytometry data, and show that our method offers improved performance.

%
\section{Background and Motivation}
\label{sec:review}

In this section, we explain the principles of flow cytometry.
We also define the statistical file matching problem in the context of flow
cytometry data, and motivate the need for an improved solution.

%
\subsection{Flow Cytometry}
\label{sec:flow_cytometry}


\begin{figure}
\begin{center}
	\includegraphics[width=0.6\linewidth]{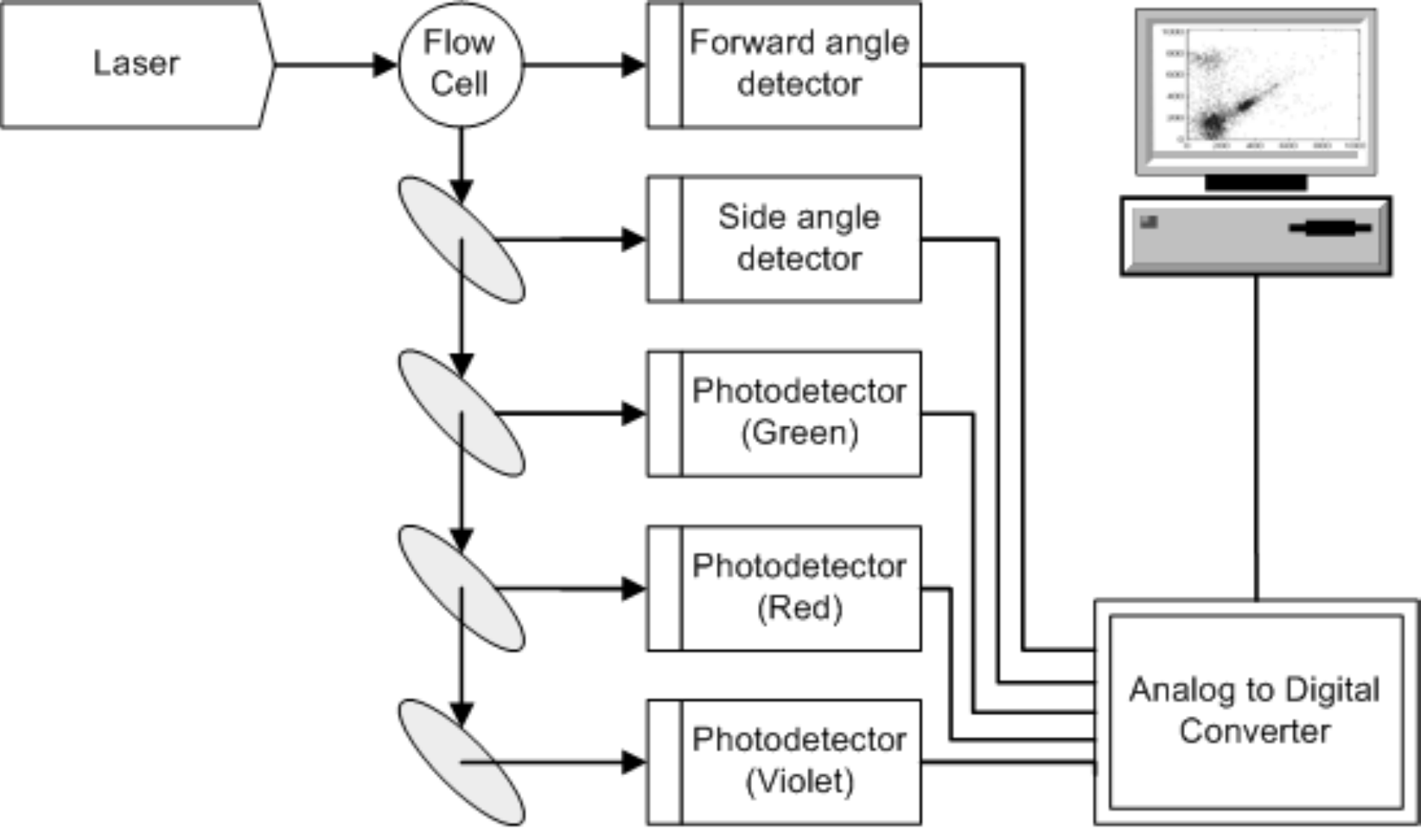}
\end{center}
%
%
\caption{
A flow cytometer system. 
As a stream of cells passes through a laser beam, the photo-detectors detect forward angle light scatter, side angle light scatter, and light emissions from fluorochromes.
Then the digitized signals are analyzed in a computer.
}
\label{fig:flow_cytometry}
\end{figure}

In flow cytometry analysis, a cell suspension is first prepared from peripheral blood, bone marrow, or lymph node.
The suspension of cells is then mixed with a solution of fluorochrome-labeled antibodies.
Typically, each antibody is labeled with a different fluorochrome.
As the stream of suspended cells passes through a focused laser beam, they either scatter or absorb the light.
If the labeled antibodies are attached to proteins of a cell, the associated fluorescent markers absorb the laser and emit light with the corresponding wavelength (color).
Then a set of photo-detectors in the line of the light beam and perpendicular to the light capture the scattered and emitted light.
The signals from the detectors are digitized and stored in a computer system.
Forward scatter (FS) and side scatter (SS) signals as well as various fluorescence signals are collected for each cell (see Fig. \ref{fig:flow_cytometry}).

In a flow cytometer that is capable of measuring $d$ attributes, called \textit{markers}, the measurements of each cell can be represented with a $d$-dimensional vector $\bx = (x^{(1)}, x^{(2)}, \cdots, x^{(d)})$ where $x^{(1)}$ is FS, $x^{(2)}$ is SS, and $x^{(3)}, \cdots, x^{(d)}$ are the fluorescent markers.
Thus, the accumulation of $N$ cells forms a $N \times d$ matrix.

The detected signals provide information about the physical and chemical properties of each cell analyzed.
FS is related to the relative size of the cell and SS is related to its internal granularity or complexity.
The fluorescence signals reflect the abundance of expressed antigens on the cell surface.
These various attributes are used for identification and quantification of cell populations.
FS and SS are always measured, while the marker combination is a part of the experimental design.

Flow cytometry data is usually analyzed using a sequence of one dimensional histograms and two or three dimensional scatter plots by choosing a subset of one, two or three markers.
The analysis typically involves manually selecting and excluding cell subpopulations, called gating, by thresholding and drawing boundaries on the scatter plots.
Clinicians routinely diagnose by visualizing the scatter plots.

Recently, some attempts have been made to analyze directly in high dimensional spaces by mathematically modeling flow cytometry data.
In \cite{boedigheimer08cytometry, chan08cytometry}, a mixture of Gaussian distributions is used to model cell populations, while a mixture of $t$-distributions with a Box-Cox transformation is used in \cite{lo08cytometry}.
A mixture of skew $t$-distributions is studied in \cite{pyne09pnas}.
The knowledge of experts is sometimes incorporated as prior information \cite{lakoumentas09bmi}.
Instead of using finite mixture models, some recent approaches proposed information preserving dimension reduction to analyze high dimensional flow cytometry data \cite{carter08ipca, carter09fine}.
However, standard techniques for multi-dimensional flow cytometry analysis are not yet established.

%
\subsection{Statistical File Matching}
\label{sec:file_matching}

\begin{figure}
\begin{center}
	\includegraphics[width=0.4\linewidth]{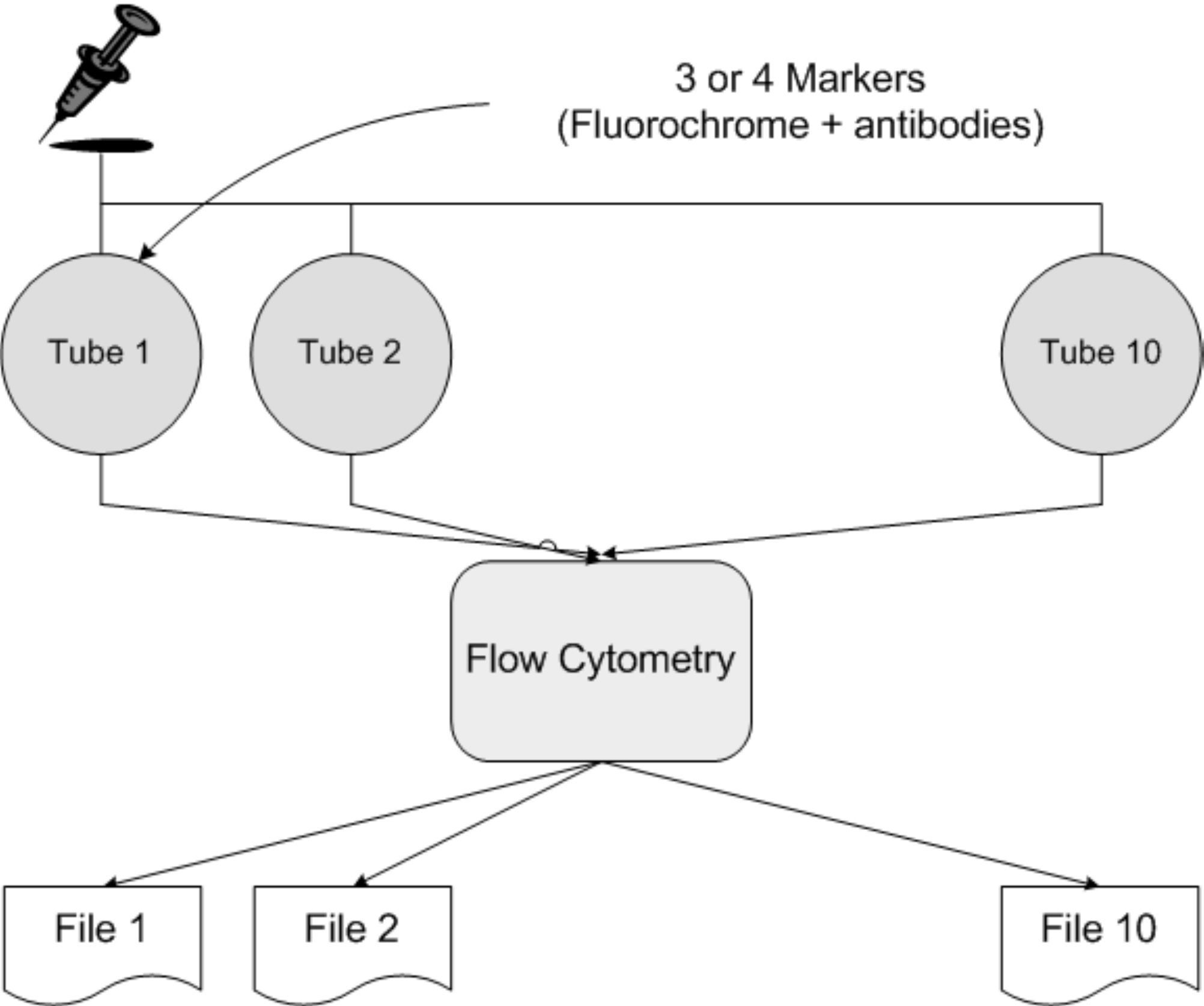}
\end{center}
%
%
\caption{Flow cytometry analysis on a large number of antibody reagents within a limited capacity of a flow cytometer.
A sample from a patient is separated into multiple tubes with which different combinations of fluorochrome labeled antibodies are stained.
Each output file contains at least two variables, FS and SS, in common as well as some variables that are specific to the file.
}
\label{fig:fc_process}
\end{figure}


The number of dimensions in flow cytometry is limited by the number of light sources and detectable fluorochrome markers, and available reagent combinations.
Even though recent innovations have enabled measuring near 20 cellular attributes, there are typically dozens or hundreds of markers of interest in a given flow cytometry experiment.
Furthermore, instruments deployed in clinical laboratories still only measure 5-7 markers simultaneously \cite{perfetto04nat}.

Being unable to simultaneously measure all markers of interest, it is common to divide a sample into several ``tubes'' and stain each tube separately with a different set of markers \cite{sanchez02leukemia}. 
In practice, partially overlapping marker combinations are used to help identify cell populations (see Fig. \ref{fig:fc_process}).
The marker combinations are designed based on which markers need to be observed together.
However, it is not always possible to anticipate all marker combinations of potential interest. 

In the sequel, we present a method that generates flow cytometry data in which all the markers of interest are available for the union of cells.
Thus, we obtain a single higher dimensional dataset beyond the current limits of instrumentation. 
Then pairs of markers that are not measured together can still be visualized through scatter plots, and methods of multidimensional analysis may be applied to the full dataset.

\begin{figure}
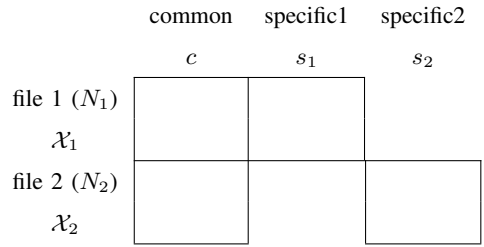

\centering
\footnotesize
\begin{tabular}{cccc}
  & common & specific1 & specific2  \\
  & $c$ & $s_1$ & $s_2$  \\
  \cline{2-3}
  file 1 ($N_1$) & \multicolumn{1}{|c|}{} & \multicolumn{1}{c|}{} &  \\
  $\cX_1$ & \multicolumn{1}{|c|}{} & \multicolumn{1}{c|}{} &  \\
  \cline{2-4}
  file 2 ($N_2$) & \multicolumn{1}{|c|}{} &  & \multicolumn{1}{|c|}{}  \\
  $\cX_2$ & \multicolumn{1}{|c|}{} &  & \multicolumn{1}{|c|}{}  \\
  \cline{2-2} \cline{4-4}
\end{tabular}
\caption{Data structure of two incomplete data files.
Two files have some overlapping variables $c$, and some variables $s_1$ and $s_2$ that are never jointly observed.
File matching combines the two files by completing the missing blocks of variables.
}
\label{fig:file_matching}
\end{figure}

This technique, called file matching, merges two nor more datasets that have some commonly observed variables as well as some variables unique to each dataset.
An exemplary two file case is drawn in Fig. \ref{fig:file_matching}.
Each unit (cell) $\bx_n$ is a vector in $\reals^d$ and belongs to one of the data files (tubes) $\cX_1$ or $\cX_2$, where each file has $N_1$ and $N_2$ units, respectively.
While variables $c$ are observed in all the units, units in $\cX_1$ have variables $s_2$ missing and units in $\cX_2$ have variables $s_1$ missing, where $s_1, s_2$, and $c$ represent specific and common variable sets.
If the observed and missing components of a unit $\bx_n$ are denoted by $o_n$ and $m_n$, then $o_n = c \cup s_1$ and $m_n = s_2$ for $\bx_n \in \cX_1$, and $o_n = c \cup s_2$ and $m_n = s_1$ for $\bx_n \in \cX_2$.

\begin{figure}[pth]
\begin{center}
	\includegraphics[width=0.5\linewidth]
	{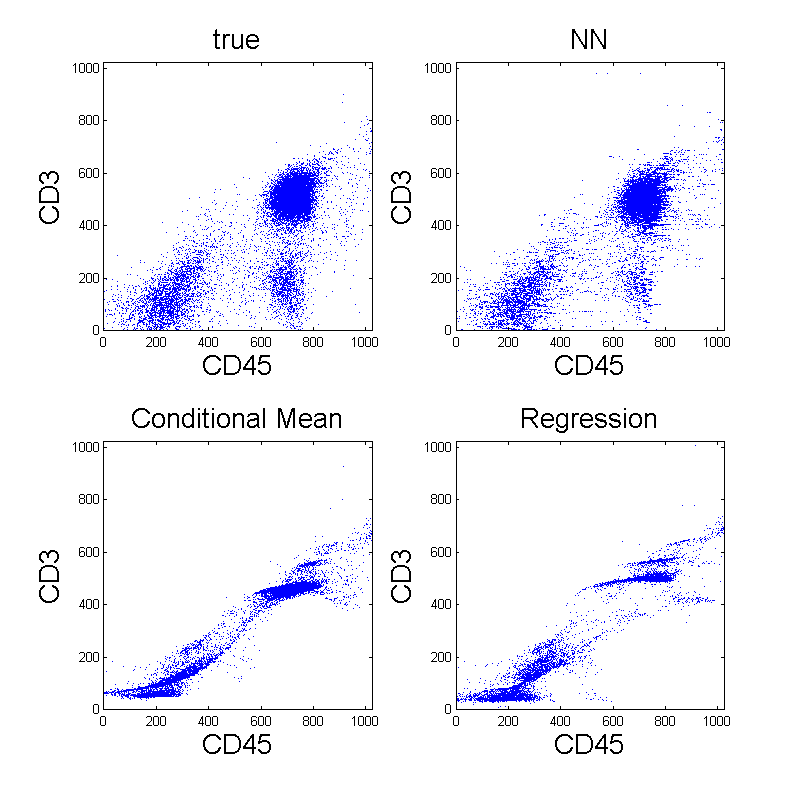}
\end{center}
%
%
\caption{Examples of imputation methods: NN, conditional mean, and regression.
The NN method relatively well preserves the distribution of imputed data, while other imputation methods such as conditional mean and regression significantly reduce the variability of data.
}
\label{fig:single_imputation}
\end{figure}

The file matching problem is a missing data problem where blocks of missing data need to be imputed.
Among imputation methods, algorithms using conditional mean or regression are most common.
As shown in Fig. \ref{fig:single_imputation}, however, these imputation algorithms tend to shrink the variance of data.
Thus, these approaches are inappropriate in flow cytometry where the shapes of cell populations are important in analysis, and the preservation of variability after file matching is highly desired.
More discussions on missing data analysis and file matching can be found in \cite{little02} and \cite{rassler02}.

\cite{pedreira08cytometry} proposed to use Nearest Neighbor (NN) imputation to match flow cytometry data files.
In their approach, missing variables of one unit, called the recipient, are imputed with observed variables from a unit in the other file, called the donor, that is most similar.
If $\bx_i$ is a unit in $\cX_1$, the missing variables are set as follows 
\begin{align*}
  \bx_i^{s_2} = \bx_j^{*\,  s_2}
  \text{ where }
  \bx_j^* = \argmin_{\bx_j \in \cX_2} \norm{\bx_i^{c} - \bx_j^{c}}_2.
\end{align*}
Note that the similarity is based on the distance in the projected space of jointly observed variables.
This algorithm is advantageous over other imputation algorithms, based on conditional mean or regression, as displayed in Fig. \ref{fig:single_imputation}.
It generally preserves the distribution of cells, while the other methods cause the variance structure to shrink toward zero.

\begin{figure}[thp]
\begin{center}
  \includegraphics[width=\linewidth]
  {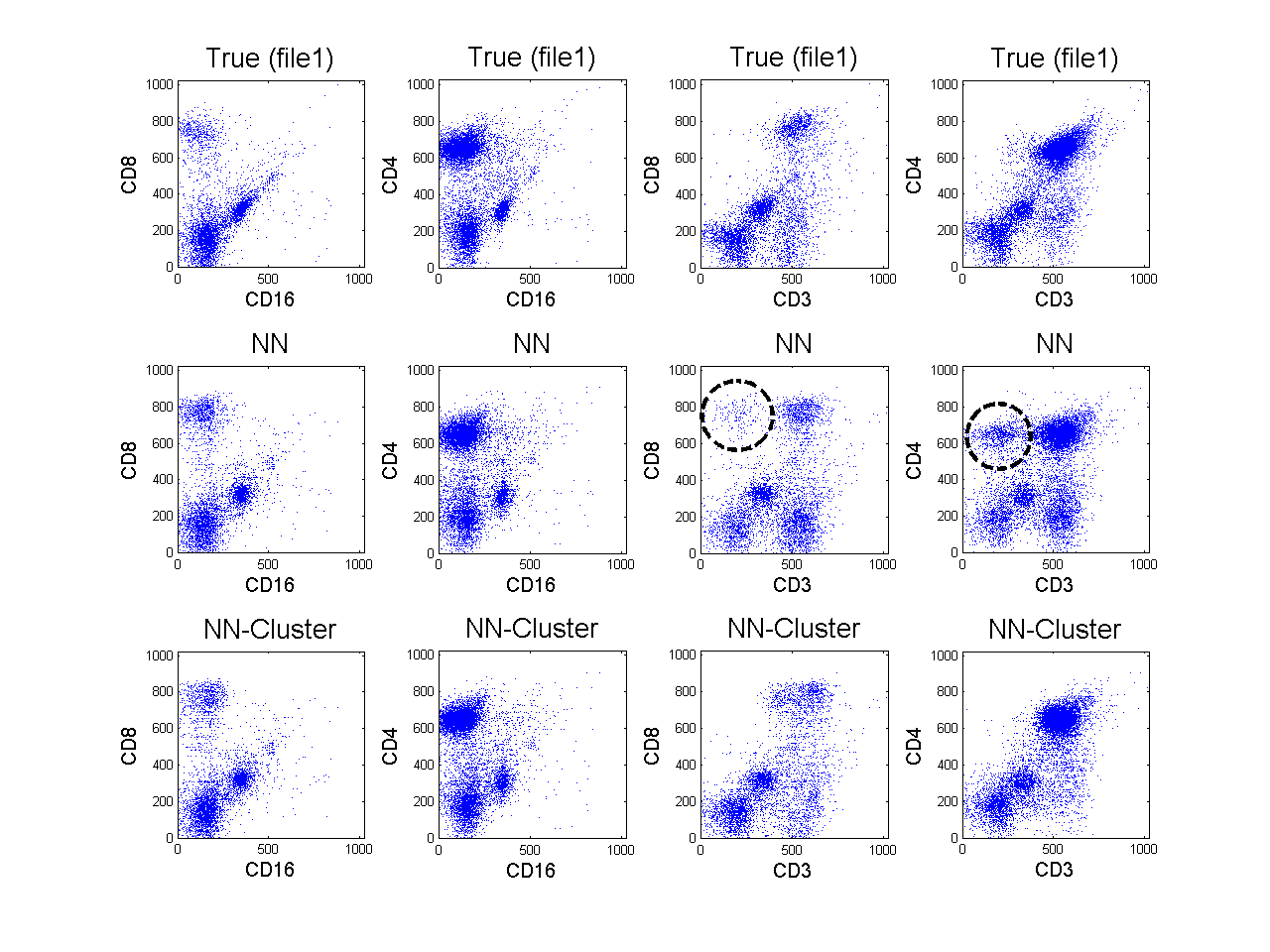}
\end{center}
%
%
\caption{
Comparison of results for two imputations methods to the ground truth cell distribution.
Figures show scatter plots on pairs of markers that are not jointly observed.
The middle row and the bottom row shows the imputation results from the NN and Cluster-based NN, respectively.
The results from the NN method show spurious clusters conspicuously in the right two panels.
The false clusters are indicated by dotted circles in CD3 vs. CD8 and CD3 vs. CD4 scatter plots.
On the other hand, the results from our proposed approach better resemble the true distribution on the top row.
}
\label{fig:exp_res}
\end{figure}

However, the NN method sometimes introduces spurious clusters into the imputed results and fails to replicate the true distribution of cell populations.
Fig. \ref{fig:exp_res} shows an example of false clusters from the NN imputation algorithm (for detailed experimental setup, see Section \ref{sec:exp}).
We present a toy example to explain why the NN imputation can fail, and to motivate our approach.

%
\subsection{Motivating Toy Example}
\label{sec:motivation}

\begin{figure}[thp]
\begin{center}
  \includegraphics[width=0.8\linewidth]
  {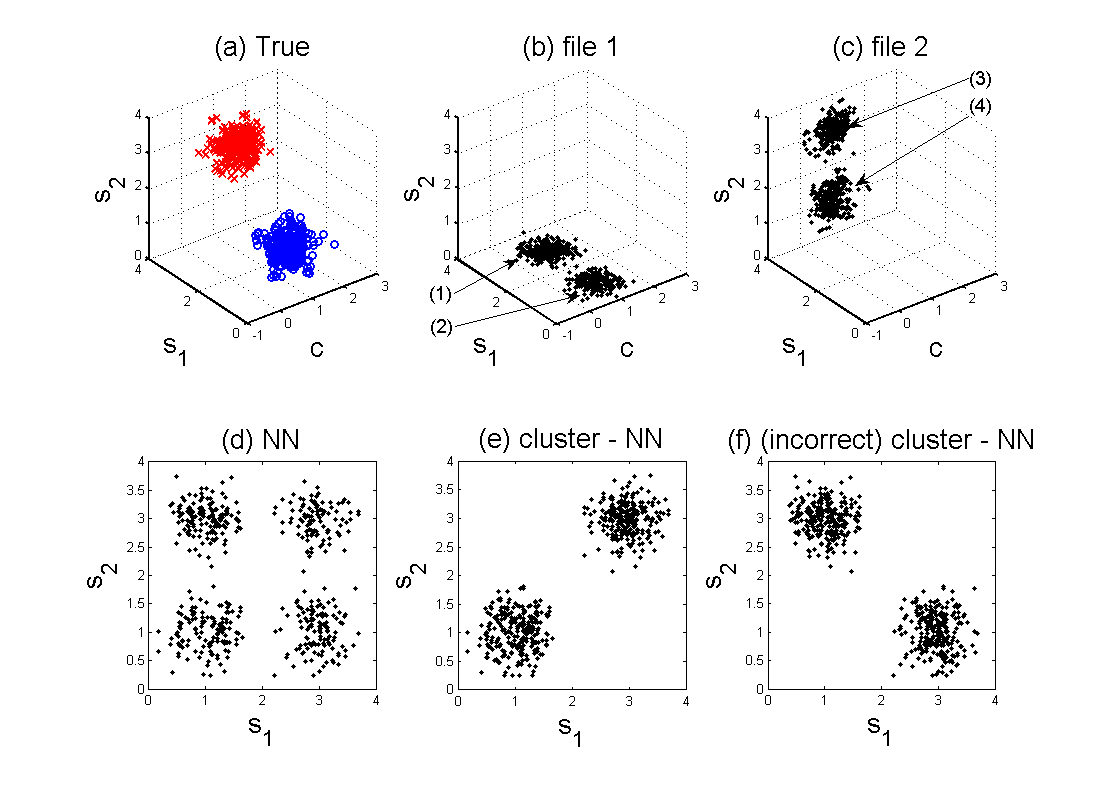}
\end{center}
%
\caption{Toy example of file matching.
Two files (b) and (c) provide partial information of data points (a) in $\reals^3$.
The variable $c$ is observed in both files while $s_1$ and $s_2$ are specific to each file.
The NN method creates false dot populations in the $s_1$ vs. $s_2$ scatter plot in (d).
On the other hand, the NN applied within the same cluster successfully replicated the true distribution.
If the cluster are incorrectly paired, however, the Cluster-NN approach fails, as in (f).
}
\label{fig:toy}
\end{figure}

Fig. \ref{fig:toy} shows a toy example dataset in $\reals^3$.
In the two data files, two of three features of these points are observed: $c$ and $s_1$ in file 1, and $c$ and $s_2$ in file 2.
Each data point belongs to one of two clusters, but its label is unavailable.

When imputing feature $s_1$ of units in file 2, the NN algorithm produces four clusters whereas there should be two, as shown in Fig. \ref{fig:toy} (d).
This is because the NN method uses only one feature, and fails to leverage the information about the joint distribution of variables that are not observed together.
On the other hand, if we can infer the cluster membership of data points, the NN imputation can be applied within the same cluster.
Hence, we search a donor from the subgroup $(1)$ for the data points in $(3)$, and likewise we search a donor from $(2)$ for the points in $(4)$ in the example.
Then the file matching result greatly improves and better replicates the true distribution as in Fig. \ref{fig:toy} (e).

In this example, as in real flow cytometry data, there is no way to infer cluster membership from the data alone, and incorrect labeling can lead to poor results (Fig. \ref{fig:toy} (f)).
Fortunately, in flow cytometry we can incorporate prior knowledge to achieve an accurate clustering.


%
\section{Cluster-based Imputation of Missing Variables}
\label{sec:imputation}

We first focus on the case of matching two files.
The case of more than two files is discussed in Section \ref{sec:discussion}.
For the present section, we assume that $\cX_1$ and $\cX_2$ have both been partitioned into $K$ clusters.
Let $\cX_1^k$ and $\cX_2^k$ denote the cells in $\cX_1$ and $\cX_2$ from the $k$th cluster, respectively.

Suppose that the data is configured as in Fig. \ref{fig:file_matching}.
In order to impute the missing variables of a unit in file 1, we locate a donor among the data points in file 2 that have the same cluster label as the recipient.
When imputing incomplete units in file 2, the roles change.
The similarity between two units is evaluated on the projected space of jointly observed variables, while constraining both units to belong to the same cluster.
Then we impute the missing variables of the recipient by patching the corresponding variables from the donor.
More specifically, for $\bx_i \in \cX_1^k$, we impute the missing variables by 
\begin{align*}
  \bx_i^{m_i} = \bx_j^{*m_i}
  \text{ where }
  \bx_j^* = \argmin_{\bx_j \in \cX_2^k} \| \bx_i^{c} - \bx_j^{c} \|_2
\end{align*}
The proposed Cluster-based NN imputation algorithm is summarized in Fig. \ref{alg:nn_cluster}.

In social applications such as survey completion, file matching is often performed on the same class such as gender, age, or county of residence.
However, this information that is used to label each unit is available in data, and the inference as in our algorithm is not necessary \cite{rassler02}.

\begin{figure}
\begin{center}
\begin{algorithmic}
  \REQUIRE two files $\cX_1$ and $\cX_2$ to be matched
  \STATE 1. Cluster the units in $\cX_1$ and $\cX_2$.
  \STATE 2. Perform NN imputation within the same cluster.
  \ENSURE statistically matched complete files $\widehat{\cX}_1$ and $\widehat{\cX}_2$
\end{algorithmic}
\end{center}
\caption{
The description of the Cluster-based NN algorithm for two files.
For two input flow cytometry data files $\cX_1$ and $\cX_2$, specific variables are imputed using NN method after clustering each cell into one of $K$ clusters.
}
\label{alg:nn_cluster}
\end{figure}

%
\section{Clustering with Missing Data}
\label{sec:missing}

To implement the above approach, we view $\cX_1$ and $\cX_2$ as a single data set and cluster its elements.
We propose a method for clustering with missing data based on a finite Gaussian mixture model.
Mixture models are common models for flow cytometry where each component corresponds to a cell type.
While non-Gaussian models might provide a better fit, there is a trade-off between estimation error and approximation error.  
More complicated models tend to be more challenging to fit.
Furthermore, even with an imperfect data model, we may still achieve an improved file matching.

Thus, clustering amounts to fitting the parameters of the mixture model.
In general, fitting such a model is ill-posed.
For example, in the toy example, there is no way to know the correct cluster
inference based solely on the data.
However, we can leverage domain knowledge to select the number of components
and initialize model parameters.

%
\subsection{Mixture of PPCA}
\label{sec:ppca}

In a mixture model framework, the probability distribution of a $d$-dimensional data vector $\bx$ takes the form
\begin{align*}
  p(\bx) = \sum_{k=1}^K \pi_k p_k(\bx)
\end{align*}
where $K$ is the number of components in the mixture and $\pi_k$ is a mixing weight.

In flow cytometry, mixture models are common models of cell subpopulations.
Mixture models with Gaussian components are common \cite{boedigheimer08cytometry, chan08cytometry, lakoumentas09bmi}, although distributions with more parameters, such as $t$-distributions or skew $t$-distri\-butions, have been proposed \cite{lo08cytometry, pyne09pnas}.
However, these models require estimating a large number of parameters, and it becomes difficult to obtain reliable estimates when the number of components or the dimensions of the data increase.
In this application, the model needs not be perfect to get improved imputation.
We adopt a probabilistic principal component analysis (PPCA) mixture model as a way to model cell populations with fewer parameters. 
Without PPCA, our experience has revealed that even a Gaussian mixture model may have too many parameters to be accurately fit.

PPCA was proposed by \cite{tipping99ppca} as a probabilistic interpretation of PCA.
While conventional PCA lacks a probabilistic formulation, PPCA specifies a generative model.
It is a latent variable model, in which a data vector is linearly related to a latent variable.
The latent variable space is generally lower dimensional than the ambient variable space, so the latent variable provides an economical representation of the data.

The PPCA model is built by specifying a conditional distribution of a data vector $\bx$ in $\reals^d$, given a latent variable $\bt$ in $\reals^q$:
\begin{align*}
  p(\bx | \bt) = \cN( \bW \bt + \bmu, \sigma^2 \bI )
\end{align*}
where $\bmu$ is a $d$-dimensional vector and $\bW$ is a $d \times q$ linear transform matrix.
The latent variable is also assumed to be Gaussian with $p(\bt) = \cN(\bnull, \bI)$.
Then the marginal distribution of $\bx$ is also Gaussian: 
\begin{align*}
  p(\bx) = \cN( \bmu, \bC )
\end{align*}
with a covariance matrix $\bC = \bW \bW^T + \sigma^2 \bI$.
The posterior distribution can be shown to be Gaussian as well: 
\begin{align*}
  p(\bt | \bx) = \cN( \bM^{-1} \bW^T (\bx - \bmu), \sigma^2 \bM^{-1} )
\end{align*}
where $\bM = \bW^T \bW + \sigma^2 \bI$ is a $q \times q$ matrix.

The PPCA mixture model is a combination of multiple PPCAs.
Each PPCA component explains local data structure or cell subpopulation.
The model is defined by the collection of each component parameters 
$\theta_k = \{ \pi_k, \bmu_k, \bW_k, \sigma^2_k \}$.
From a flow cytometry dataset $\cX = \{ \bx_1, \cdots, \bx_N \}$, an EM algorithm can learn the mixture model by iteratively computing these parameters.
More details on the PPCA mixture and the EM algorithm are explained in \cite{tipping99mppca}

The mixture of PPCA offers a way of controlling the number of parameters to be estimated without completely sacrificing the flexibility of model.
In mixture model framework, a more common choice is the standard Gaussian mixture model.
In the Gaussian mixture model, however, each Gaussian component requires $d(d+1)/2$ covariance parameters to be estimated if a full covariance matrix is used.
Thus, as the data dimension increases, more data points are needed for reliable estimation of those parameters.
The number of parameters can be reduced by constraining the covariance matrix to be isotropic or diagonal.
These are too restrictive, however, since an isotropic or diagonal covariance makes the Gaussian component spherical or, respectively, elliptical aligned along the data axes; hence, the correlation structure between variables cannot be captured.
On the other hand, the PPCA mixture model lies between those two extremes, and allows to control the number of parameters by specifying $q$, the dimension of the latent variable.

%
\subsection{Mixture of PPCA with Missing Data}
\label{sec:ppca_missing}

Even though our file matching problem has a particular pattern of missing variables, we develop a more general algorithm that allows for an arbitrary pattern of missing variables.  
Our development assumes values are ``missing at random,'' meaning that whether a variable is missing or not, is independent of its value \cite{little02}. 
Our algorithm may be viewed as an extension of the algorithm of \cite{ghahramani94nips} to PPCA, or the algorithm of \cite{tipping99mppca} to data with missing values.

Denoting the observed and missing components by $o_n$ and $m_n$, each data point can be divided $\bx_n = \left( \bx_n^{o_n}, \bx_n^{m_n} \right)$.
In a missing data problem, a set of partial observations $\{ \bx_1^{o_1}, \cdots, \bx_N^{o_N} \}$ is given. 
Similar to EM algorithms for Gaussian mixture models, we introduce indicator variables $\bz_n$.
One and only one entry of $\bz_n$ is nonzero, and $z_{nk}=1$ indicates that the $k$th component is responsible for generating $\bx_n$.
We also include the missing components $\bx_n^{m_n}$ and the set of latent variables $\bt_{nk}$ for each component to form the complete data $(\bx_n^o, \bx_n^m, \bt_{nk}, \bz_n)$ for $n=1, \cdots, N$ and $k=1, \cdots, K$.

We derive an iterative EM algorithm for the PPCA mixture model with missing data.
The key difference from the EM algorithm for completely observed data is that the conditional expectation is taken with respect to $\bx^o$ as opposed to $\bx$ in the expectation steps.

To develop an EM algorithm, we employ and extend the two step procedure as described in \cite{tipping99mppca}.
In the first stage of the algorithm, the component weights $\pi_k$ and the component center $\bmu_k$ are updated: 
\begin{align}
  \widehat{\pi}_k 
  = & \frac{1}{N} \sum_n \cond{z_{nk}},  \\
  \widehat{\bmu}_{k}
  = & \frac{\sum_n \cond{z_{nk}} 
    \Big[ \begin{array}{c} \bx_n^{o_n} \\ \cond{\bx_n^{m_n}} \end{array} \Big]}
    {\sum_n \cond{z_{nk}}}
\end{align}
where $\cond{z_{nk}} = P(z_{nk}=1 | \bx_n^{o_n})$ is the responsibility of mixture component $k$ for generating the unit $\bx_n$, and $\cond{\bx_n^{m_n}} = \E [ \bx_n^{m_n} | z_{nk}=1, \bx_n^{o_n} ]$ is the conditional expectation.
Note that we are not assuming the vectors in the bracket are stackable.
This notation can be replaced by the true component ordering without difficulty.

In the second stage, we update $\bW_k$ and $\sigma^2_k$: 
\begin{align}
  \widehat{\bW}_{k}
  = & \bS_{k} \bW_{k} 
    (\sigma^2_{k} \bI + \bM_{k}^{-1} \bW_{k}^T \bS_{k} \bW_{k})^{-1},  \\
  \widehat{\sigma}^2_{k}
  = & \frac{1}{d} \tr 
    \left( \bS_{k} - \bS_{k} \bW_{k} \bM_{k}^{-1} \widehat{\bW}_{k}^T \right)
\end{align}
from local covariance matrix $\bS_k$: 
\begin{align*}
  \bS_k = \frac{1}{N \widehat{\pi}_{k}} 
    \sum_n \cond{z_{nk}} 
    \cond{(\bx_n - \widehat{\bmu}_k) (\bx_n - \widehat{\bmu}_k)^T}.
\end{align*}
These update rules boil down to the update rules for completely observed data when there are no missing variables.
We derive the EM algorithm in detail in Appendix \ref{sec:mppca_em_missing}.

After model parameters are estimated, the observations are divided into groups according to their posterior distribution: 
\begin{align*}
  \argmax_{k=1, \cdots K} p \, (z_{nk}=1 | \bx_n^{o_n}), 
\end{align*}
so each unit (cell) is classified into one of $K$ cell populations.
Note that this posterior probability is computed in the E-step.

%
\subsection{Domain Knowledge and Initialization of EM algorithm}
\label{sec:prior}

\begin{figure}
\centering
\footnotesize
\begin{tabular}{l|l|l}
  \multicolumn{2}{c|}{Cell Type} & CD markers  \\
  \hline \hline
  \multicolumn{2}{c|}{granulocyte} & CD45+, CD15+  \\
  \hline
  \multicolumn{2}{c|}{monocyte} & CD45+, CD14+  \\
  \hline
  \multirow{4}{*}{lymphocyte} & helper T cell & CD45+, CD3+  \\
   & cytotoxic T cell & CD45+, CD3+, CD8+  \\
   & B cell & CD45+, CD19+ or CD45+, CD20+  \\
   & NK cell & CD16+, CD56+, CD3-  \\
\end{tabular}
\caption{
Types of white blood cells. 
Each cell type is characterized by a set of expressed CD markers. 
The cluster of differentiation (CD) markers are commonly used to identify cell surface molecules on white blood cells.
The `$+/-$' signs indicate whether a certain cell type has corresponding antigens on the cell surface.
}
\label{fig:cell_type}
\end{figure}

In file matching of flow cytometry data, domain knowledge is critical.
First, as explained above, the incompletely observed data is insufficient to determine the correct cluster labeling.
Second, the initial conditions of the EM algorithm affect its performance and convergence rate.
Domain knowledge allows us to choose the number of components, and to initialize the algorithm so that it converges to the best local maximum.

In flow cytometry, from the design of fluorochrome marker combinations and the knowledge about the blood sample composition, we can anticipate certain properties of cell subpopulations. 
For example, Fig. \ref{fig:cell_type} summarizes white blood cell types and their characteristic cluster of differentiation (CD) marker expressions.
That these are six cell types suggests choosing $K=6$ when analyzing white blood cells.

The CD markers indicated are commonly used in flow cytometry to identify cell surface molecules on leukocytes \cite{zola05blood}.
However, this information is qualitative, and needs to be quantified.

To achieve this, we use one dimensional histograms.
In a histogram, two large peaks are generally expected depending on the expression level of the corresponding CD marker.
If a cell subpopulation expresses a CD marker, denoted by `$+$', then it forms a peak on the right side of the histogram.
On the other hand, if a cell population does not express the marker, denoted by `$-$', then a peak can be found on the left side of the histogram.
We use the locations of the peaks to quantify the expression levels.

These quantified values can be combined with the CD marker expression levels of each cell type to specify the initial cluster centers.
Thus, each component of $\bmu_k$ of a certain cell type is initialized by either the positive quantity or the negative quantity from the histogram.
In our implementation, these are set manually based on visual inspection of histograms.
Then we initialize the mixture model parameters $\{ \pi_k, \bmu_k, \bW_k, \sigma^2_k \}$ as described in Fig. \ref{alg:em_init}.

\begin{figure}
\begin{center}
\begin{algorithmic}
  \REQUIRE $\cX_1$, $\cX_2$ data files ; 
    $K$ the number of components ; 
    $q$ the dimension of latent variable space ; 
    $\bmu_k$ for initial component mean.
  \FOR{$k=1$ to $K$}
    \STATE 1. using distance $\| \bx_n^{o_n} - \bmu_k^{o_n} \|$, find the set 
      of data points $\cX^k$ whose nearest component mean is $\bmu_k$
    \STATE 2. initialize a covariance matrix $\bC_k$ with random entries
    \STATE 3. replace submatrices of $\bC_k$ with sample covariance of data 
      points in $\cX^k$
    \STATE 4. make $\bC_k$ positive definite by enforcing the eigenvalues to 
      be positive
    \STATE 5. set $\pi_k = \frac{|\cX^k|}{N_1 + N_2}$
    \STATE 6. set $\bW_k$ with the $q$ principal eigenvectors of $\bC_k$
    \STATE 7. set $\sigma^2_k$ with the average of remaining eigenvalues of 
      $\bC_k$
  \ENDFOR
  \ENSURE $\{ \pi_k$, $\bmu_k$, $\bW_k$, $\sigma^2_k \}$ for $k=1, \cdots, K$
\end{algorithmic}
\end{center}
\caption{
Parameter initialization of an EM algorithm for missing data.
Cell populations are partitioned into $K$ groups based on the distance to each component center.
The component weight $\pi_k$ is initialized according to the size of each partition.
From the covariance matrix estimate $bC_k$, parameters $bW_k$ and $\sigma^2_k$ are initialized by taking eigen-decomposition.
}
\label{alg:em_init}
\end{figure}

\begin{figure}
\centering
\begin{tabular}{*{4}{c}}
  & $\ c \, $ & $s_1$ & $s_2$  \\
  \cline{2-4}
  $c$ & \multicolumn{1}{|c|}{} &  & \multicolumn{1}{|c|}{}  \\
  \cline{2-4}
  $s_1$ & \multicolumn{1}{|c|}{} &  & \multicolumn{1}{|c}{}   \\
  \cline{2-4}
  $s_2$ & \multicolumn{1}{|c|}{} &  & \multicolumn{1}{|c|}{}  \\
  \cline{2-2} \cline{4-4}
\end{tabular}
\caption{
Structure of covariance matrix $\bC$.
The sub-matrices $\bC_k^{s_1, s_2}$ and $\bC_k^{s_2, s_1}$ cannot be estimated from a sample covariance matrix because these variables are never jointly observed.
}
\label{fig:cov_mtx}
\end{figure}

An important issue in file matching arises from the covariance matrix.
When data is completely observed, a common way of initialization of a covariance matrix is using a sample covariance matrix.
In the case of file matching, however, it cannot be evaluated since some sets of variables are never jointly observed (see Fig. \ref{fig:cov_mtx}).
We chose to build a covariance matrix $\bC_k$ from variable to variable with sample covariances.
For example, we can set $\bC_k^{c, s_1}$ with the sample covariance for variables $c$ and $s_1$ based on cases for which both variables $c$ and $s_1$ are present.
On the other hand, the submatrix $\bC_k^{s_1, s_2}$ cannot be built based on the observation.
In our implementation, we set the submatrix $\bC_k^{s_1, s_2}$ with arbitrary values.
However, the resulting matrix may not be positive definite.
Thus, $\bC_k$ is made positive definite by replacing negative eigenvalues with a small positive value.
Once a covariance matrix $\bC_k$ is obtained, we can initialize $\bW_k$ and $\sigma^2_k$ by taking eigen-decomposition of $\bC_k$.

%
\section{Experiments and Results}
\label{sec:exp}

We apply the proposed file matching technique to real flow cytometry datasets, and present experimental results.

Three flow cytometry datasets are prepared from lymph node samples of three patients.
These datasets were provided by the Department of Pathology at the University of Michigan.
The measurements are of different sizes and have seven attributes: FS, SS, CD56, CD16, CD3, CD8, and CD4.
Each dataset is randomly permuted ten times and divided into two data files and a separate evaluation set.
In Fig. \ref{fig:exp_sets}, the cell counts of the two files and the held-out set are denoted $N_1$, $N_2$, and $N_e$, respectively.
Two attributes from each file are made hidden to construct hypothetical files with missing data.
Thus, CD16 and CD3 are available only in file 1, and CD8 and CD4 are available only in file 2, while FS, SS, and CD56 are common.
The pattern of the constructed data files is illustrated in Fig. \ref{fig:exp_file} where the blocks of missing variables are left blank.

\begin{figure}[tbp]
\begin{center}
\footnotesize
\begin{tabular}{c|rrr}
  ID & $N_1$ & $N_2$ & $N_e$  \\
  \hline
  Patient1 & 10000 & 10000 & 5223  \\
  Patient2 &  7000 &  7000 & 4408  \\
  Patient3 &  3000 &  3000 & 3190 
\end{tabular}
\end{center}
\caption{Three flow cytometry datasets from three different patients.
Each dataset is divided into two data files and an evaluation set.
$N_1$ and $N_2$ denote the size of two data files and $N_e$ is the size of evaluation set.
}
\label{fig:exp_sets}
\end{figure}

\begin{figure}[tbp]
\begin{center}
\footnotesize
\begin{tabular}{lccccccc}
   & FS & SS & CD56 & CD16 & CD3 & CD8 & CD4  \\
  \cline{2-4} \cline{5-6}
  file 1 & \multicolumn{3}{|c|}{} & \multicolumn{2}{c|}{} & 
     \multicolumn{2}{c}{}  \\
  \cline{2-4} \cline{5-6} \cline{7-8}
  file 2 & \multicolumn{3}{|c|}{} & 
     \multicolumn{2}{c|}{} & \multicolumn{2}{c|}{}  \\
  \cline{2-4} \cline{7-8}
\end{tabular}
\end{center}
\caption{File structure used in the experiment. 
FS, SS, and CD56 are common in both files, and a pair of CD markers are observed in only one of the files.
The blank blocks correspond to the unobserved variables.
The blocks in file 1 are matrices with $N_1$ rows, and the blocks in file 2 are matrices with $N_2$ rows.
}
\label{fig:exp_file}
\end{figure}

\begin{figure}[tbp]
\begin{center}
	\includegraphics[width=\linewidth]{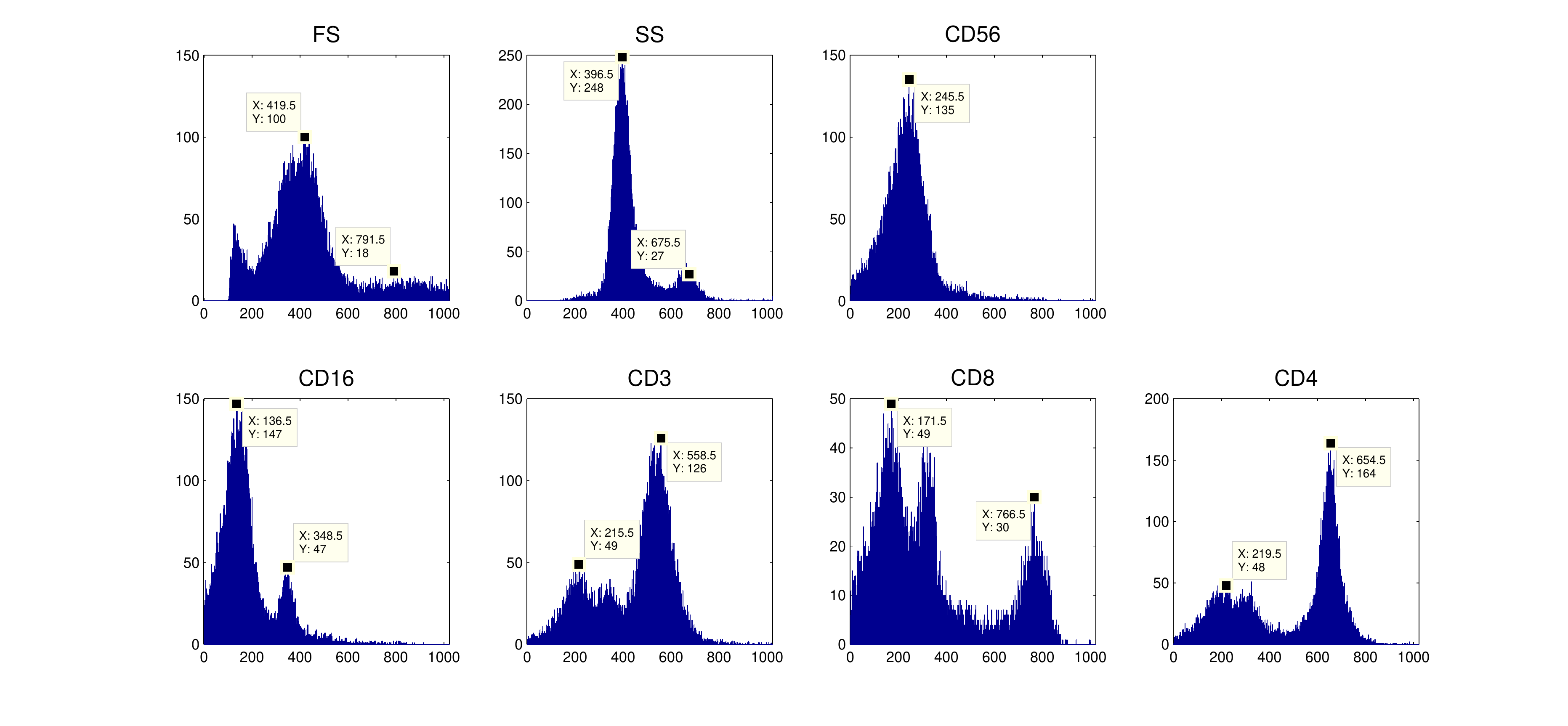}
\end{center}
%
%
\caption{
Histogram of each marker in the dataset.
The peaks are hand-picked and are indicated in each panels.
}
\label{fig:exp_hist}
\end{figure}

For each white blood cell type, its expected marker expressions (CD markers), relative size (FS), and relative granularity (SS) are presented in Fig. \ref{fig:exp_cell}.
Because it is from a lymph node sample, the majority of cell population is lymphocytes, while the most common white blood cells in a human body are granulocytes.
The `$+/-$' signs indicate whether a certain cell type expresses the markers or not.
For example, helper T cells express both CD3 and CD4 but not others.
This qualitative knowledge is quantified with the help of single dimensional histograms as explained in Section \ref{sec:prior}.
Two dominant peaks are picked from each histogram and their corresponding measurement values are set to the positive and negative expression levels.
Fig. \ref{fig:exp_hist} and Fig. \ref{fig:exp_hist_quan} summarize this histogram analysis.

\begin{figure}[tbp]
\begin{center}
\footnotesize
\begin{tabular}{l|ccccccc}
  Cell type & FS & SS & CD56 & CD16 & CD3 & CD8 & CD4  \\
  \hline
  granulocyte         & $+$ & $+$ & $-$ & $+$ & $-$ & $-$ & $-$  \\
  monocyte            & $+$ & $-$ & $-$ & $+$ & $-$ & $-$ & $-$  \\
  helper T cell       & $-$ & $-$ & $-$ & $-$ & $+$ & $-$ & $+$  \\
  cytotoxic T cell    & $-$ & $-$ & $-$ & $-$ & $+$ & $+$ & $-$  \\
  B lymphocyte        & $-$ & $-$ & $-$ & $-$ & $-$ & $-$ & $-$  \\
  Natural Killer cell & $-$ & $-$ & $+$ & $+$ & $-$ & $-$ & $-$  
\end{tabular}
\end{center}
\caption{
Cell types in the dataset and their corresponding marker expressions.
`+' or `-' indicates whether a certain cell type expresses the CD marker or not.
}
\label{fig:exp_cell}
\end{figure}

\begin{figure}[tbp]
\centering
\footnotesize
\begin{tabular}{c|ccccccc}
 & FS & SS & CD56 & CD16 & CD3 & CD8 & CD4  \\
\hline
$+$ & 800 & 680 & 500 & 350 & 550 & 750 & 650  \\
$-$ & 400 & 400 & 240 & 130 & 200 & 170 & 200  \\
\end{tabular}
\caption{
The positive and negative expression levels are summarized.
}
\label{fig:exp_hist_quan}
\end{figure}

Two incomplete data files are completed following the procedure as described in Fig. \ref{alg:nn_cluster}.
A mixture of PPCA is fitted with six components because six cell types are expected from this dataset.
The latent variable dimension of each PPCA component is fixed to two.

The synthesized data after file matching is displayed in Fig. \ref{fig:exp_res}.
The figure shows scatter plots of specific variables: CD16, CD3, CD4, and CD8.
Note that these marker pairs are not jointly observed from the two incomplete data files.
The imputation results from the NN and the Cluster-based NN methods are compared in the figure.
For reference, scatter plots from the original complete dataset (ground truth) are also presented.
As can be seen, the results from the Cluster-based NN are far more similar to the true distributions.
On the other hand, the results from the NN method generates spurious clusters in the CD3-CD8 and CD3-CD4 scatter plots.
In Fig. \ref{fig:exp_res}, these false clusters are indicated.

%
\subsection{Evaluation method}
\label{sec:evaluation}

To quantitatively evaluate the previous results, we use Kullback-Leibler (KL) divergence.
The KL divergence between two distribution $f(\bx)$ and $g(\bx)$ is defined by 
\begin{align*}
  \kl{g}{f} 
  = \E_g \left[ \log g - \log f \, \right].
\end{align*}
Let $f$ denote a true distribution responsible for the observations, and $g$ denote its estimate.

The KL divergence is not symmetric, so $\kl{f}{g}$ and $\kl{g}{f}$ have different meanings.
For a given distribution $f$, a distribution $g$ minimizes $\kl{f}{g}$ when $g$ takes nonzero values in the region where $f$ takes nonzero values; hence, it overestimates the support of $f$.
On the other hand, $\kl{g}{f}$ is minimized for $g$ that is close to zero in the region where $f$ is near zero.
A distribution $g$ that minimizes $\kl{g}{f}$ tends to have smaller support. 
Therefore, $\kl{g}{f}$ is a better evaluation method for detecting spurious clusters in an estimate.

Then the empirical estimate of the KL divergence is evaluated by
\begin{align*}
  \kl{g}{f}
  \approx \kl{\widehat{g}}{\widehat{f}}
  \approx \frac{1}{N_e} \sum_{n=1}^{N_e} 
    \left[ \log \widehat{g} \, (\widehat{\bx}_n) - \log \widehat{f} \, (\widehat{\bx}_n) \, \right] .
\end{align*}
where the distributions $f$ and $g$ are replaced by their corresponding density estimates, and the expectation is approximated by a finite sum over imputed results $\widehat{\bx}_n$ on the held-out validation set of size $N_e$.

We used kernel density estimation on the ground truth data and the imputed data for $\widehat{f}$ and $\widehat{g}$, respectively.
The KL divergences are computed for ten random permutations, and their averages and standard errors are reported in Fig. \ref{fig:eval_kl}.
As can be seen, the KL divergences from Cluster-based NN are substantially smaller than those from NN.
Therefore, the Cluster-based NN yields a better replication of true distribution.

\begin{figure}
\centering
\footnotesize
\begin{tabular}{c|cc|cc}
  ID  & NN (file 1) & Cluster-NN (file 1) & NN (file 2) & Cluster-NN (file 2) \\
  \hline
  Patient1  &  2.90 $\pm$ 0.05  &  1.55 $\pm$ 0.05  &  
            2.66 $\pm$ 0.03  &  1.12 $\pm$ 0.04  \\
  Patient2  &  4.54 $\pm$ 0.07  &  1.22 $\pm$ 0.03  &  
            4.12 $\pm$ 0.08  &  0.92 $\pm$ 0.03  \\
  Patient3  &  4.46 $\pm$ 0.10  &  2.40 $\pm$ 0.11  &  
            4.18 $\pm$ 0.11  &  2.30 $\pm$ 0.07
\end{tabular}
\caption{
The KL divergences are computed for ten permutations of each flow cytometry dataset.
The averages and standard errors are reported in the table.
For both the NN and Cluster-based NN algorithm, the file matching results are evaluated.
The KL divergences of Cluster-based NN are closer to zero than those of NN.
Thus the results from Cluster-based NN better replicated the true distribution.
}
\label{fig:eval_kl}
\end{figure}

%
\section{Discussion}
\label{sec:discussion}

%

In this paper, we demonstrated the use of a cluster-based nearest neighbor imputation method for file matching in flow cytometry data.
We applied the proposed algorithm on real flow cytometry data to generate a dataset of higher dimensions by merging two data files of lower dimensions.
The resulting matched file can be used for visualization and high-dimensional analysis of cellular attributes.

While the presented imputation method focused on the case of two files, it can be generalized to more than two files.
For each missing component of a recipient cell, we can find a donor cell among files that have the component of interest.
We envision two extensions of the clustering-based imputation method.
The first is training a PPCA mixture model on all the data files.
This approach involves the entire data points for model fitting.
The second method considers a pair of files at a time.
In this approach, we first select a donor file in which the missing component of the recipient file is available.
Then we apply method of this paper to the pair of files.
This approach involves smaller number of data points in training, but mixture models of smaller dimensions need to be fitted multiple times.
After training of a mixture model and clustering of each cell, the similarity between cells can be computed.
The Euclidean distance on the projected space of commonly observed variables can be used to find the similarity under the constraint that both units should have the same cluster label.
The missing components are then imputed from the donor.

Future research directions include finding ways of automatic prior information extraction.
The construction of covariance matrices from incomplete dataset in the initialization of the EM algorithm is also an interesting problem.
We expect that better covariance structure estimation will be helpful for better replication of non-symmetric and non-elliptic cell populations in the imputed results.

A limitation of this work is that it has only been validated on lymphocyte data, where, for certain marker combinations, cell types tend to form relatively well-defined clusters.  
However, for other samples and marker combinations, clusters may be more elongated or less well-defined due to cells being at different stages of physiologic development.  
Future work may also consider more flexible models for clustering such data, and associated inference algorithms.

%
\appendix

%
\section{Derivation of EM Algorithm for Mixture of PPCA model with missing data}
\label{sec:mppca_em_missing}

Suppose that we are given an incomplete observation set.
We can divide each unit $\bx_n$ as $\bx_n = \left[ \begin{array}{cc} \bx_n^{o_n} \\ \bx_n^{m_n} \end{array} \right]$ by separating the observed components and the missing components.
Note that we do not assume that the observed variables are first, and the notation can be replaced by the actual ordering of components without difficulty.

In the PPCA mixture model, the probability distribution of $\bx$ is 
\begin{align*}
  p(\bx) = \sum_{k=1}^K \pi_k p(\bx | k).
\end{align*}
where $K$ is the number of components in the mixture and $\pi_k$ is a mixing weight corresponding to the component density $p(\bx | k)$.
We estimate the set of unknown parameters $\theta = \{ \pi_k, \bmu_k, \bW_k, \sigma^2_k \}$ using an EM algorithm from the partial observations $\{ \bx_1^{o_1}, \cdots, \bx_N^{o_N} \}$.

To develop an EM algorithm, we introduce indicator variables $\bz_n = (z_{n1}, \cdots, z_{nK})$ for $n=1, \cdots, N$.
One and only one entry of $\bz_n$ is nonzero, and $z_{nk}=1$ indicates that the $k$th component is responsible for generating $\bx_n$.
We also include a set of the latent variables $\bt_{nk}$ for each component, and missing variables $\bx_n^{m_n}$ to form the complete data 
$(\bx_n^{o_n}, \bx_n^{m_n}, \bt_{nk}, \bz_n)$ for $n=1, \cdots, N$ and $k=1, \cdots, K$.
Then the corresponding complete data likelihood function has the form: 
\begin{align*}
  \eq{L}_C
  = & \sum_n \sum_k z_{nk} \ln \left[ \pi_k p(\bx_n, \bt_{nk}) \right]  \\
  = & \sum_n \sum_k z_{nk} \bigg[ \ln \pi_k - \frac{d}{2} \ln \sigma^2_{k} 
    - \frac{1}{2 \sigma_{k}^2} 
      \tr \left[ (\bx_n - \bmu_{k})(\bx_n - \bmu_{k})^T \right]  \notag  \\
    & + \frac{1}{\sigma^2_{k}} 
      \tr \left[ (\bx_n - \bmu_{k}) \bt_{nk}^T \bW_{k}^T \right]
    - \frac{1}{2 \sigma^2_{k}} 
      \tr \left[ \bW_{k}^T \bW_{k} \bt_{nk} \bt_{nk}^T \right]
    \bigg]
\end{align*}
%
%
where terms independent of the parameters are not included in the second equality.
Instead of developing an EM algorithm directly on this likelihood function $\eq{L}_C$, we extend the strategy in \cite{tipping99mppca} and build a two-stage EM algorithm, where each stage is a two-step process.

In the first stage of the two stage EM algorithm, we update the component weight $\pi_k$ and the component mean $\bmu_k$.
We form a complete data log-likelihood function with the component indicator variables $\bz_n$ and missing variables $\bx_n^m$ while ignoring the latent variables $\bt_{nk}$.
Then we have the following likelihood function: 
\begin{align*}
  \eq{L}_1 
  = & \sum_{n=1}^N \sum_{k=1}^K z_{nk} \ln [ \pi_k p(\bx_n^{o_n}, \bx_n^{m_n} | k) ]  \\
  = & \sum_n \sum_k z_{nk} \left[ \ln \pi_k - \frac{1}{2} \ln | \bC_k | 
    - \frac{1}{2} \tr \left[ \bC_k^{-1} 
      (\bx_n - \bmu_k)(\bx_n - \bmu_k)^T \right]
    \right]
\end{align*}
%
%
where terms unrelated to the model parameters are omitted in the second line.
We take the conditional expectation with respect to $p(\bz_n, \bx_n^{m_n} | \bx_n^{o_n})$.
Since the conditional probability factorizes as 
\begin{align*}
  p(\bz_n, \bx_n^{m_n} | \bx_n^{o_n}) 
  = p(\bz_n | \bx_n^{o_n}) p(\bx_n^{m_n} | \bz_n, \bx_n^{o_n}), 
\end{align*}
the next conditional expectations follow 
%
\begin{align*}
  \cond{z_{nk}} 
  = & p(k | \bx_n^{o_n}) 
  = \frac{\pi_k p(\bx_n^{o_n} | k)}
    {\sum_{k'} \pi_{k'} p(\bx_n^{o_n} | k')},  \\
  \cond{z_{nk} \bx_n^{m_n}}
  = & \cond{z_{nk}} \cond{\bx_n^{m_n}},  \\
  \cond{\bx_n^{m_n}}
  = & \bmu_k^{m_n} + \bC_k^{m_n o_n} \bC_k^{{o_n o_n}^{-1}} (\bx_n^{o_n} - \bmu_k^{o_n}),  \\
  \cond{z_{nk} \bx_n^{m_n} \bx_n^{{m_n}^T}}
  = & \cond{z_{nk}} \cond{\bx_n^{m_n} \bx_n^{{m_n}^T}},  \\
  \cond{\bx_n^{m_n} \bx_n^{{m_n}^T}}
  = & \bC_k^{m_n m_n} - \bC_k^{m_n o_n} \bC_k^{{o_n o_n}^{-1}} \bC_k^{o_n m_n}
    + \cond{\bx_n^{m_n}} \cond{\bx_n^{{m_n}^T}}
\end{align*}
where $\cond{\cdot}$ denote the conditional expectation.
Maximizing $\cond{\eq{L}_1}$ with respect to $\pi_k$, using a Lagrange multiplier, and with respect to $\mu_k$ give the parameter updates
\begin{align}
  \widehat{\pi}_k 
  = & \frac{1}{N} \sum_n \cond{z_{nk}},  
  \label{eq:pi_k}  \\
  \widehat{\bmu}_{k}
  = & \frac{\sum_n \cond{z_{nk}} 
    \Big[ \begin{array}{c} \bx_n^{o_n} \\ \cond{\bx_n^{m_n}} \end{array} \Big]}
    {\sum_n \cond{z_{nk}}}.
    \label{eq:mu_k}
\end{align}

In the second stage, we include the latent variable $\bt_{nk}$ as well to formulate the complete data log-likelihood function.
The new values of $\widehat{\pi}_k$ and $\widehat{\bmu}_{k}$ are used in this step to compute sufficient statistics.
Taking the conditional expectation on $\eq{L}_C$ with respect to $p(\bz_n,  \bt_{nk}, \bx_n^{m_n} | \bx_n^{o_n})$, we have 
\begin{align*}
  \cond{\eq{L}_C}
  = & \sum_n \sum_k \cond{z_{nk}} \bigg[ \ln \widehat{\pi}_k 
    - \frac{d}{2} \ln \sigma^2_{k} 
    - \frac{1}{2 \sigma_{k}^2} 
      \tr \left[ \cond{(\bx_n - \widehat{\bmu}_{k})
      (\bx_n - \widehat{\bmu}_{k})^T} \right]  \notag  \\
    & + \frac{1}{\sigma^2_{k}} 
      \tr \left[ \cond{(\bx_n - \widehat{\bmu}_{k}) \bt_{nk}^T} 
      \bW_{k}^T \right]
    - \frac{1}{2 \sigma^2_{k}} 
      \tr \left[ \bW_{k}^T \bW_{k} \cond{\bt_{nk} \bt_{nk}^T} \right]
  \bigg].
\end{align*}
%
%
Since the the conditional probability factorizes
\begin{align*} 
  p(\bz_n, \bt_{nk}, \bx_n^{m_n} | \bx_n^{o_n}) 
  = p(\bz_n | \bx_n^{o_n}) p(\bx_n^{m_n} | \bz_n, \bx_n^{o_n}) 
    p(\bt_{nk} | \bz_n, \bx_n^{o_n}, \bx_n^{m_n}), 
\end{align*}
%
we can evaluate the conditional expectations as follows : 
\begin{align*}
  \cond{(\bx_n - \widehat{\bmu}_{k})(\bx_n - \widehat{\bmu}_{k})^T}
  = & \left( \left[ \begin{array}{c} 
      \bx_n^{o_n} \\ 
      \cond{\bx_n^{m_n}} \end{array} \right] - \widehat{\bmu}_k \right)
    \left( \left[ \begin{array}{c} 
      \bx_n^{o_n} \\ 
      \cond{\bx_n^{m_n}} \end{array} \right] - \widehat{\bmu}_k \right)^T
    + \left[ \begin{array}{cc} 
        \bnull & \bnull \\ 
        \bnull & \bQ_{nk} 
      \end{array} \right],  \\
  Q_{nk} 
  = & \bC_k^{m_n m_n} - \bC_k^{m_n o_n} \bC_k^{{o_n o_n}^{-1}} \bC_k^{o_n m_n},  \\
  \cond{\bt_{nk}}
  = & \bM_k^{-1} \bW_k^T (\bx_n - \widehat{\bmu}_k),  \\
  \cond{(\bx_n - \widehat{\bmu}_{k}) \bt_{nk}^T}
  = & \cond{(\bx_n - \widehat{\bmu}_{k})(\bx_n - \widehat{\bmu}_{k})^T} 
    \bW_k \bM_k^{-1},  \\
  \cond{\bt_{nk} \bt_{nk}^T} 
  = & \sigma^2_k \bM_k^{-1} 
    + \bM_k^{-1} \bW_k^T 
      \cond{(\bx_n - \widehat{\bmu}_{k})(\bx_n - \widehat{\bmu}_{k})^T} 
      \bW_k \bM_k^{-1}.
\end{align*}
%
%
Remember that the $q \times q$ matrix $\bM_k = \bW_k^T \bW_k + \sigma^2_k \bI$.
Then the maximization of $\cond{\eq{L}_C}$ with respect to $\bW_k$ and $\sigma^2_k$ leads to the parameter updates, 
\begin{align}
  \widehat{\bW}_{k}
  = & \left[ \sum_n \cond{z_{nk}} 
    \cond{(\bx_n - \widehat{\bmu}_{k}) \bt_{nk}^T} \right]
    \left[ \sum_n \cond{z_{nk}} \cond{\bt_{nk} \bt_{nk}^T} \right]^{-1},  \\
  \widehat{\sigma}^2_k
  = & \frac{1}{d \sum_n \cond{z_{nk}}}
    \bigg[ \sum_n \cond{z_{nk}} 
      \tr \left[ \cond{(\bx_n - \widehat{\bmu}_{k})
      (\bx_n - \widehat{\bmu}_{k})^T} \right]  \\
    & - 2 \sum_n \cond{z_{nk}} 
      \tr \left[ \cond{(\bx_n - \widehat{\bmu}_{k}) \bt_{nk}^T} 
      \bW_{k}^T \right]  \notag  \\
    & + \sum_n \cond{z_{nk}} 
      \tr \left[ \bW_{k}^T \bW_{k} \cond{\bt_{nk} \bt_{nk}^T} \right] 
    \bigg]  \notag  .
\end{align}
%
%
Substituting the conditional expectations simplifies the M-step equations 
\begin{align}
  \widehat{\bW}_{k}
  = & \bS_{k} \bW_{k} 
    (\sigma^2_{k} \bI + \bM_{k}^{-1} \bW_{k}^T \bS_{k} \bW_{k})^{-1},  
    \label{eq:W_k}  \\
  \widehat{\sigma}^2_{k}
  = & \frac{1}{d} \tr 
    \left( \bS_{k} - \bS_{k} \bW_{k} \bM_{k}^{-1} \widehat{\bW}_{k}^T \right)
    \label{eq:sigma_k}
\end{align}
where
\begin{align*}
  \bS_k = \frac{1}{N \widehat{\pi}_{k}} 
    \sum_n \cond{z_{nk}} 
    \cond{(\bx_n - \widehat{\bmu}_k) (\bx_n - \widehat{\bmu}_k)^T}.
\end{align*}

Each iteration of the EM algorithm updates the set of old parameters $\{\pi_k, \bmu_k, \bW_k, \sigma^2_k \}$ with the set of new parameters $\{\widehat{\pi}_k, \widehat{\bmu}_k, \widehat{\bW}_k, \widehat{\sigma}^2_k \}$ as in \eqref{eq:pi_k}, \eqref{eq:mu_k}, \eqref{eq:W_k}, and \eqref{eq:sigma_k}.
The algorithm terminates when the value of the log-likelihood function no longer changes.

%




\bibliographystyle{IEEEtran}
\bibliography{refs}

%





\end{document}